\documentclass[preprint,journal]{vgtc}

\preprinttext{\footnotesize
To appear in IEEE Transactions on Visualization and Computer Graphics.}

\manuscriptnote{%
\copyright\ 2026 IEEE. Personal use of this material is permitted.
Permission from IEEE must be obtained for all other uses, in any current or future media,
including reprinting/republishing this material for advertising or promotional purposes,
creating new collective works, for resale or redistribution to servers or lists, or reuse
of any copyrighted component of this work in other works.}

\graphicspath{{figures/}{pictures/}{images/}{./}}

\usepackage{times}

\usepackage{booktabs}
\usepackage{amsmath}
\usepackage{mathptmx}
\usepackage{xcolor}
\usepackage{graphicx}
\usepackage{pifont}
\usepackage{amsfonts}
\newcommand{\cmark}{\ding{51}}
\newcommand{\xmark}{\ding{55}}
\newcommand{\pmark}{\ensuremath{\triangle}} 

\onlineid{0}
\vgtccategory{Research}

\title{AtlasLC: Fast Codec-Ready Compression of Object-Centric 3D Gaussian Splatting}

\author{%
  \authororcid{ByungHyun Kim}{0009-0005-4891-3487},
  \authororcid{Jinwoo Jeon}{0009-0008-7649-8913}, and
  \authororcid{Woontack Woo}{0000-0002-5501-4421}
}

\authorfooter{
  \item ByungHyun Kim is with KAIST UVR Lab.
  E-mail: kimbh00@kaist.ac.kr.
  \item Jinwoo Jeon is with KAIST KI-ITC ARRC.
  E-mail: zkrkwlek@kaist.ac.kr.
  \item Woontack Woo is with KAIST UVR Lab.
  E-mail: wwoo@kaist.ac.kr.
}

\teaser{
\centering
\includegraphics[
  width=\linewidth,
  alt={Seven-stage overview of AtlasLC for object-centric 3D Gaussian Splatting compression. The input Gaussians are transformed into a single-pass conditional coordinate system and divided into coarse local-competition cells. Locally redundant Gaussians are pruned while support is retained across the object. The surviving Gaussians are then deterministically and collision-freely packed into a two-dimensional codec atlas, which is compressed with a standard image or video codec for XR deployment.}
]{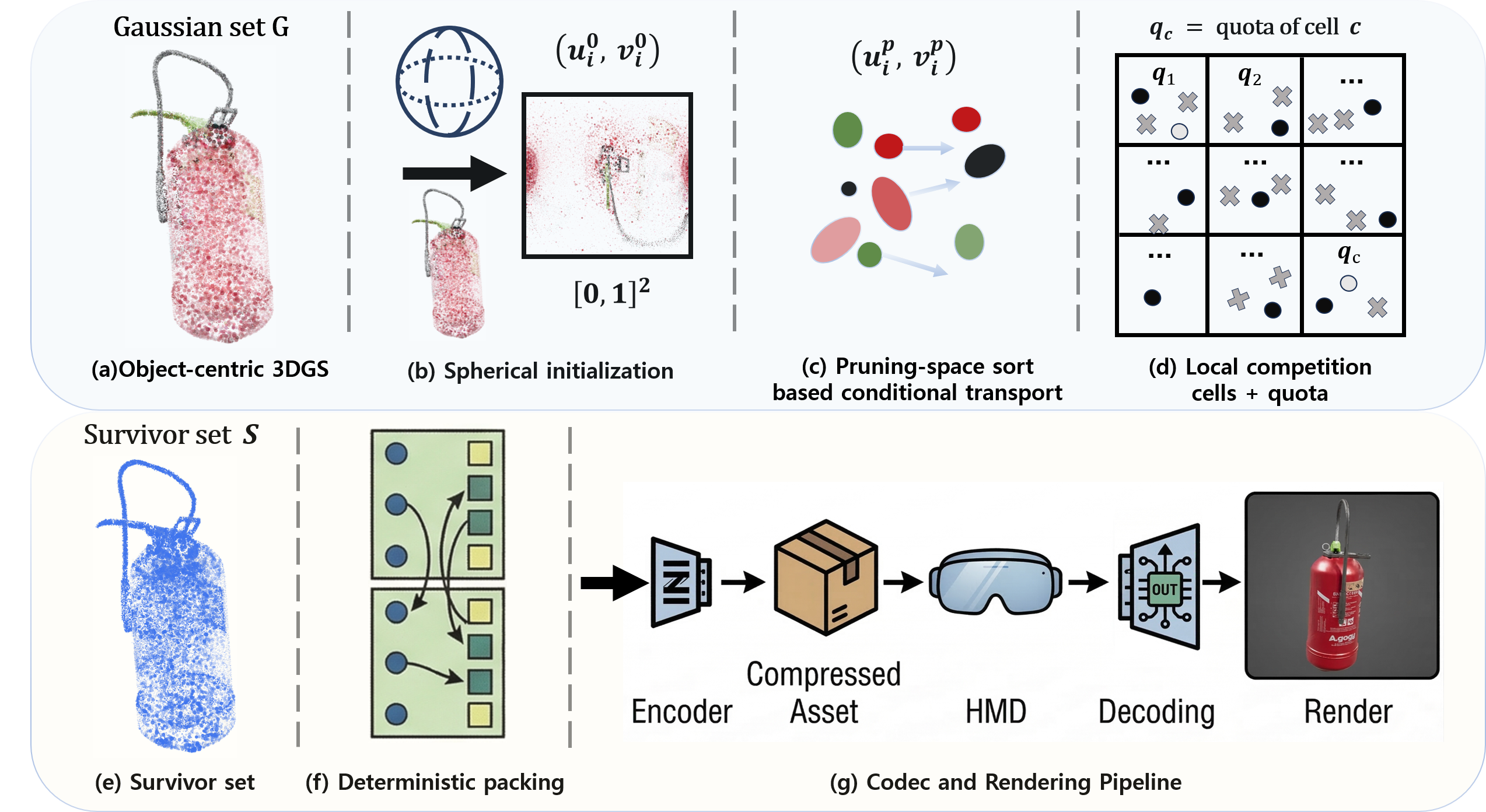}\caption{
\textbf{Pipeline of AtlasLC for object-centric 3DGS compression.}
AtlasLC first builds a \emph{single-pass} sort-based conditional coordinate system over the input Gaussians, then applies \emph{local-competition pruning} within coarse coordinate cells to remove locally redundant splats while preserving object-wide foreground support (a\,--\,d).
The retained Gaussians are finally \emph{deterministically packed} into a codec-friendly atlas for standard image/video compression (e\,--\,g).
The resulting pipeline is training-free, lightweight, and designed for deployable XR object assets, balancing compactness, fast preparation, decoding efficiency, and geometry preservation.
}
\label{fig:teaser}
}

\abstract{
3D Gaussian Splatting (3DGS) enables photorealistic novel-view synthesis with real-time rendering, but deploying compressed \emph{object-centric} 3DGS in XR requires more than image-space rate--distortion. In practical XR asset pipelines, reusable objects are repeatedly packaged, transmitted, decoded, and instantiated, making asset-preparation cost, codec compatibility, decoding latency, and preservation of depth and silhouette cues first-class concerns. Existing 3DGS compression methods are largely developed for scene-scale captures and often rely on heavy layout generation or aggressive global pruning, assumptions that transfer poorly to semantically concentrated foreground objects. We present AtlasLC, a source-free, training-free compression pipeline for object-centric 3DGS that operates directly on released Gaussian assets, without original images, camera poses, or per-asset optimization. AtlasLC couples local-competition pruning with deterministic atlas packing to remove the mapping/remapping bottleneck while preserving object-wide foreground support; a lightweight single-pass sort-based conditional transport is used as a shared coordinate backbone for these stages. Across the evaluated assets, AtlasLC reduces atlas-preparation time
by up to 25$\times$ and end-to-end compression time by up to
5$\times$, while offering a favorable deployment-aware balance of
payload, decode latency, runtime FPS, and 3D geometry relative to
the evaluated compressed baselines. Relative to similarly compact structured baselines, it uses about 6--8\% fewer bits while maintaining comparable perceptual and geometric quality. These results show that object-centric 3DGS compression should be optimized for a \emph{deployment-aware operating point} enabling scalable XR asset libraries.
}


\begin{document}

\firstsection{Introduction}
\maketitle
3D Gaussian Splatting (3DGS) has rapidly emerged as a strong representation for photorealistic novel-view synthesis, combining high visual fidelity with real-time rendering \cite{Kerbl2023_3DGS}. As 3DGS moves from reconstruction benchmarks toward deployable XR and mobile content, however, the bottlenecks shift from rendering quality alone to end-to-end deployability \cite{Li2025_RadianceFieldsXR,Franke2025_VRSplatting,Du2026_MobileGS}. In practical XR systems, especially on untethered head-mounted and low-power mobile devices, immersive content is increasingly fetched from a server or edge system and rendered under tight wireless throughput, latency, reliability, and power constraints \cite{Luo2023_3GPPXR,Alnajim2023_XRTraffic,Gao2025_XRgo}. This pressure is already well recognized for volumetric 6DoF media and, more recently, for 3DGS itself, where scene delivery and adaptive streaming have become explicit research problems \cite{Liu2022_Vues,Tsai2025_L3GS,Shi2025_LapisGS}. Compression therefore matters not only for storage and transmission, but also for startup delay, decoding latency, and whether an asset can be fetched and rendered quickly enough for interactive XR use \cite{Li2025_RadianceFieldsXR,Tsai2025_L3GS,Gao2025_XRgo,Du2026_MobileGS}.

In many XR pipelines, the deployable unit is not a full captured scene but a reusable \emph{object asset}: for example, a product preview, tool, or furniture item loaded on demand in a mixed-reality experience \cite{Dogan2024_XRObjects}. This shift is reflected in internet-scale object repositories \cite{Deitke2023_Objaverse}, object-centric 3D benchmarks \cite{Liu2025_uCO3D}, digital-twin catalogs \cite{Dong2025_DTC}, and generative asset pipelines \cite{Yang2025_GenAssets}. In such settings, assets are repeatedly packaged, compressed, transmitted, decoded, cached, updated, and redeployed across authoring and runtime stages \cite{Friston2017_3DRepo4Unity,Tsai2025_L3GS}. For object libraries, compression is therefore not merely a final archival step. The turnaround of \emph{asset preparation itself}---including preprocessing, packaging, encoding, and decoding---directly affects how quickly assets can be published, refreshed, and redeployed across heterogeneous XR systems.

We focus on a source-free, post-hoc setting where compression operates on a released 3DGS asset only, without original images, camera poses, or per-asset fine-tuning. This framing aligns with recent post-hoc 3DGS compression settings that operate on trained Gaussian assets rather than on the original capture data \cite{Xie2024_MesonGS,Tian2025_FlexGaussian,Bagdasarian2025_3DGSzip}. Recent source-free or optimization-free methods such as MesonGS, FlexGaussian, and FCGS further suggest that \emph{compression-time cost itself} is a practical concern, not just the final bitrate \cite{Xie2024_MesonGS,Tian2025_FlexGaussian,Chen2025_FCGS}. In the XR object-asset regime, preprocessing cost, codec compatibility, decoding latency, and stable repeated deployment matter alongside rate--distortion \cite{Li2025_RadianceFieldsXR,Franke2025_VRSplatting,Du2026_MobileGS,Xie2024_MesonGS,Tian2025_FlexGaussian,Chen2025_FCGS}.

For such assets, preserving geometry-related cues is also essential. Geometry-aware Gaussian variants already show that depth fidelity and surface consistency matter beyond appearance alone \cite{Huang2024_2DGS,Zhang2026_RaDeGS,Guedon2024_SuGaR}. In AR/MR and XR, reliable depth ordering supports occlusion \cite{Macedo2023_OcclusionHandlingAR,Walton2017_AccurateOcclusionMR}; stable silhouettes and boundaries matter for alignment, selection, and interaction \cite{Hellmuth2021_MobileARPositioning,Lilija2019_OccludedInteraction,Tang2020_GrabAR,Dogan2024_XRObjects}; and depth-faithful geometry proxies are useful for collision handling, editing, physics, and simulation-oriented workflows \cite{Guedon2024_SuGaR,Guedon2024_GaussianFrosting,Li2025_GaussianUDF,Ma2025_MaGS,Liu2025_ArtGS}. A method that reduces bytes by destabilizing silhouettes or depth is therefore less useful for deployment than one that preserves these cues under a similar bitrate.

Most prior 3DGS compression methods were developed under scene-centric assumptions and are commonly evaluated through scene-level rate--distortion \cite{Bagdasarian2025_3DGSzip,Yang2024_GGSCBenchmark}. Existing approaches either impose structure on unordered Gaussians for codec compatibility or reduce storage through pruning, quantization, and learned compaction \cite{Morgenstern2024_SelfOrganizingGaussianGrids,Rai2025_UVGS,Lee2025_CodecGS,Yang2025_LGSCV,Kim2025_VQHEVCHybrid3DGS,Niedermayr2024_Compressed3DGS,Lee2024_Compact3DGS,Papantonakis2024_ReducingMemoryFootprint3DGS,Fan2024_LightGaussian,Wang2024_ContextGS,Chen2024_HAC,Navaneet2024_CompGSVQ,Cao2025_LightweightPredictiveGS,Xie2024_MesonGS,Tian2025_FlexGaussian,Chen2025_FCGS}. For object-centric assets, however, two mismatches matter: layout generation can dominate asset-preparation cost, and removable redundancy is often local rather than globally disposable. Because much of the Gaussian budget lies near the visible object surface, aggressive global pruning can damage silhouettes and depth support even when bitrate is reduced.
These observations suggest that \textbf{object-centric 3DGS compression should be treated as a distinct XR deployment problem}. The goal is to reduce bytes under an \textbf{asset-only, training-free} constraint while also minimizing \textbf{asset-preparation and total compression pipeline time} and preserving the foreground support needed for silhouettes, depth, and interaction \cite{Xie2024_MesonGS,Tian2025_FlexGaussian,Chen2025_FCGS,Huang2024_2DGS,Guedon2024_SuGaR,Zhang2026_RaDeGS,Liu2025_ArtGS,Guo2025_ArticulatedGS,Ma2025_MaGS}. This favors methods that are \emph{training-free}, \emph{fast to prepare}, \emph{HMD codec-compatible}, \emph{deterministic across runs}, and \emph{conservative about local foreground support} \cite{Xie2024_MesonGS,Chen2025_FCGS,Tian2025_FlexGaussian,Lee2025_CodecGS,Kim2025_VQHEVCHybrid3DGS,Zhang2024_LP3DGS,Hanson2025_PUP3DGS}. To address this need, we present \textbf{AtlasLC}, a \textbf{training-free} compression pipeline for \textbf{object-centric 3D Gaussian Splatting} that operates directly on a released 3DGS asset, without original captures, camera metadata, fine-tuning, or per-asset learned adaptation. AtlasLC is centered on two deployment-critical operations: \textbf{local-competition pruning}, which removes redundancy within coarse neighborhoods while preserving object-wide support, and \textbf{deterministic atlas packing}, which converts the retained Gaussians into a stable, collision-free, codec-friendly representation for standard image/video codecs. A lightweight \emph{single-pass} sort-based conditional transport is used only as a shared coordinate backbone for these two stages, avoiding iterative transport optimization and any additional remapping stage.

\begin{figure}[t]
\centering
\begin{minipage}[t]{0.49\columnwidth}
    \centering
    \includegraphics[
  width=\linewidth,
  alt={Panel a. Line chart of compressed payload versus PSNR for AtlasLC, PLAS, LGSCV, UVGS, and FlexGaussian across multiple codec operating points. Within the common payload range of AtlasLC, PLAS, and LGSCV, AtlasLC has slightly lower PSNR than the structured baselines while remaining close in image-space fidelity.}
]{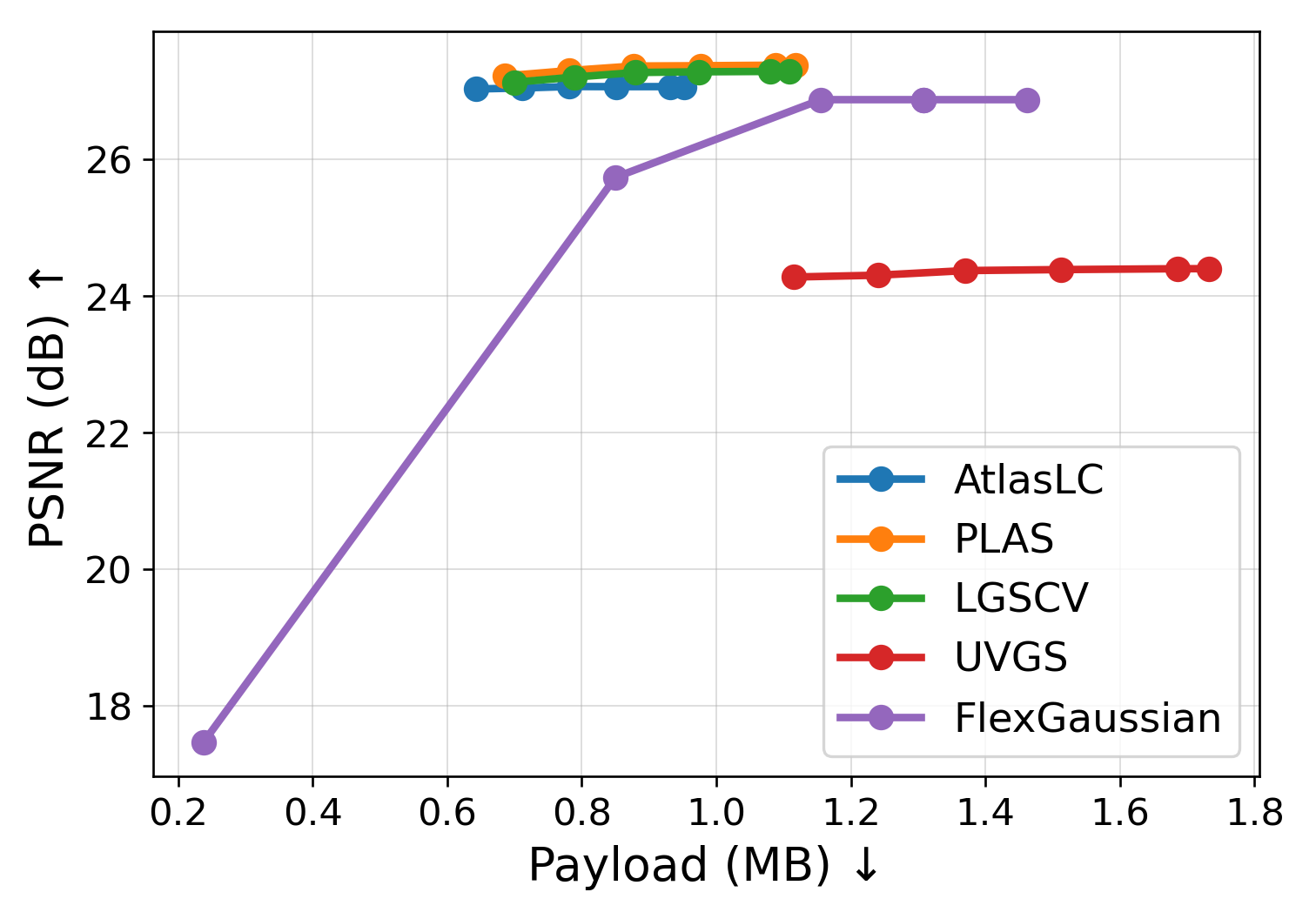}
    \vspace{-1.5mm}

    {\scriptsize \textbf{(a)} Payload--PSNR.}
\end{minipage}\hfill
\begin{minipage}[t]{0.49\columnwidth}
    \centering
\includegraphics[
  width=\linewidth,
  alt={Panel b. Line chart of compressed payload versus end-to-end compression pipeline time for AtlasLC, PLAS, LGSCV, UVGS, and FlexGaussian. AtlasLC maintains substantially lower total pipeline time than PLAS and LGSCV across their shared payload range. The timing curves are not strictly monotonic because measured runtime varies across codec operating points.}
]{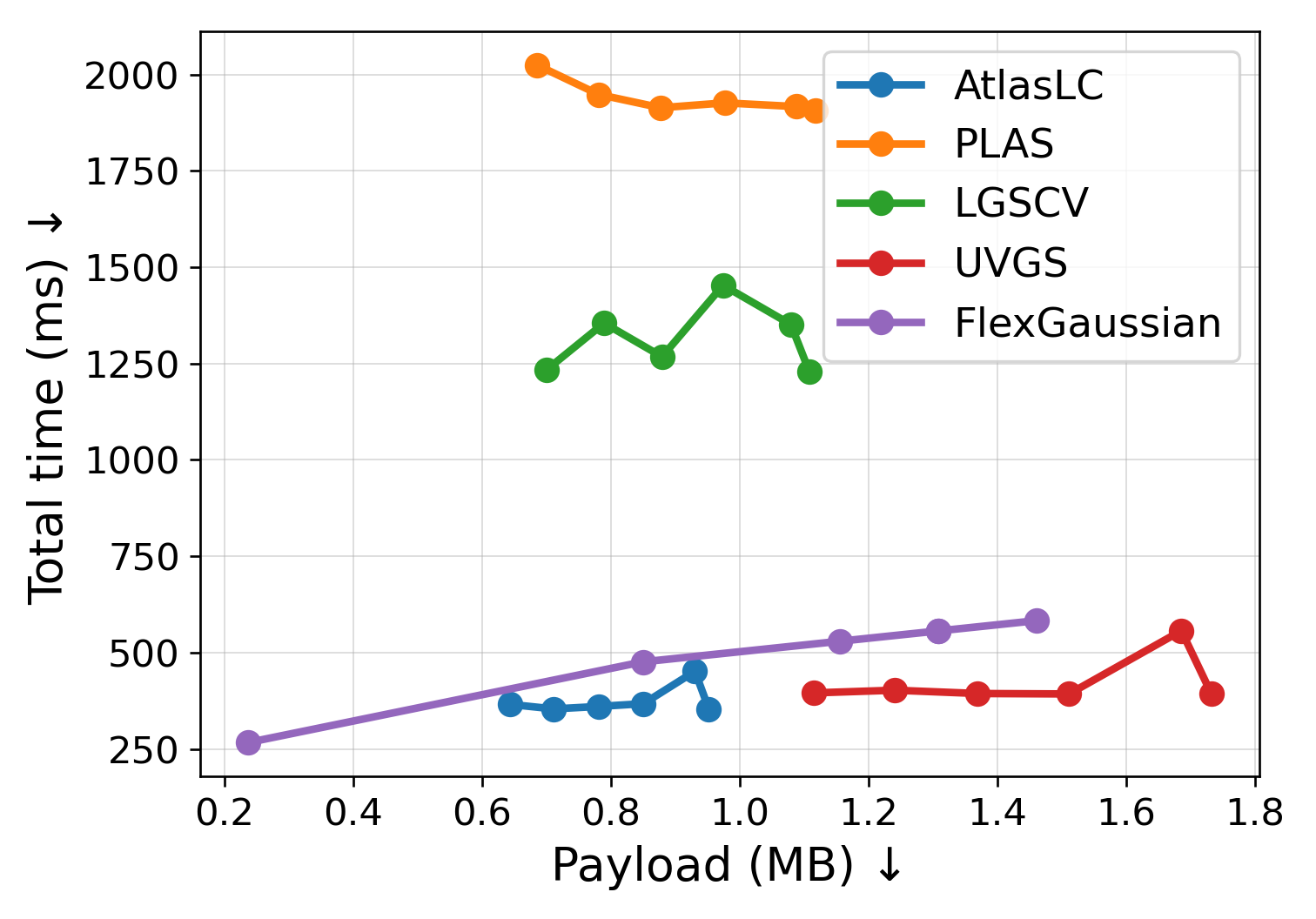}
    \vspace{-1.5mm}

    {\scriptsize \textbf{(b)} Payload--total time.}
\end{minipage}

\vspace{0.4mm}

\begin{minipage}[t]{0.49\columnwidth}
    \centering
\includegraphics[
  width=\linewidth,
  alt={Panel c. Line chart of compressed payload versus three-dimensional F1 precision for AtlasLC, PLAS, LGSCV, UVGS, and FlexGaussian. Within the shared payload range, AtlasLC generally achieves higher precision than PLAS and LGSCV, indicating that a larger fraction of its retained points lies close to the ground-truth object surface.}
]{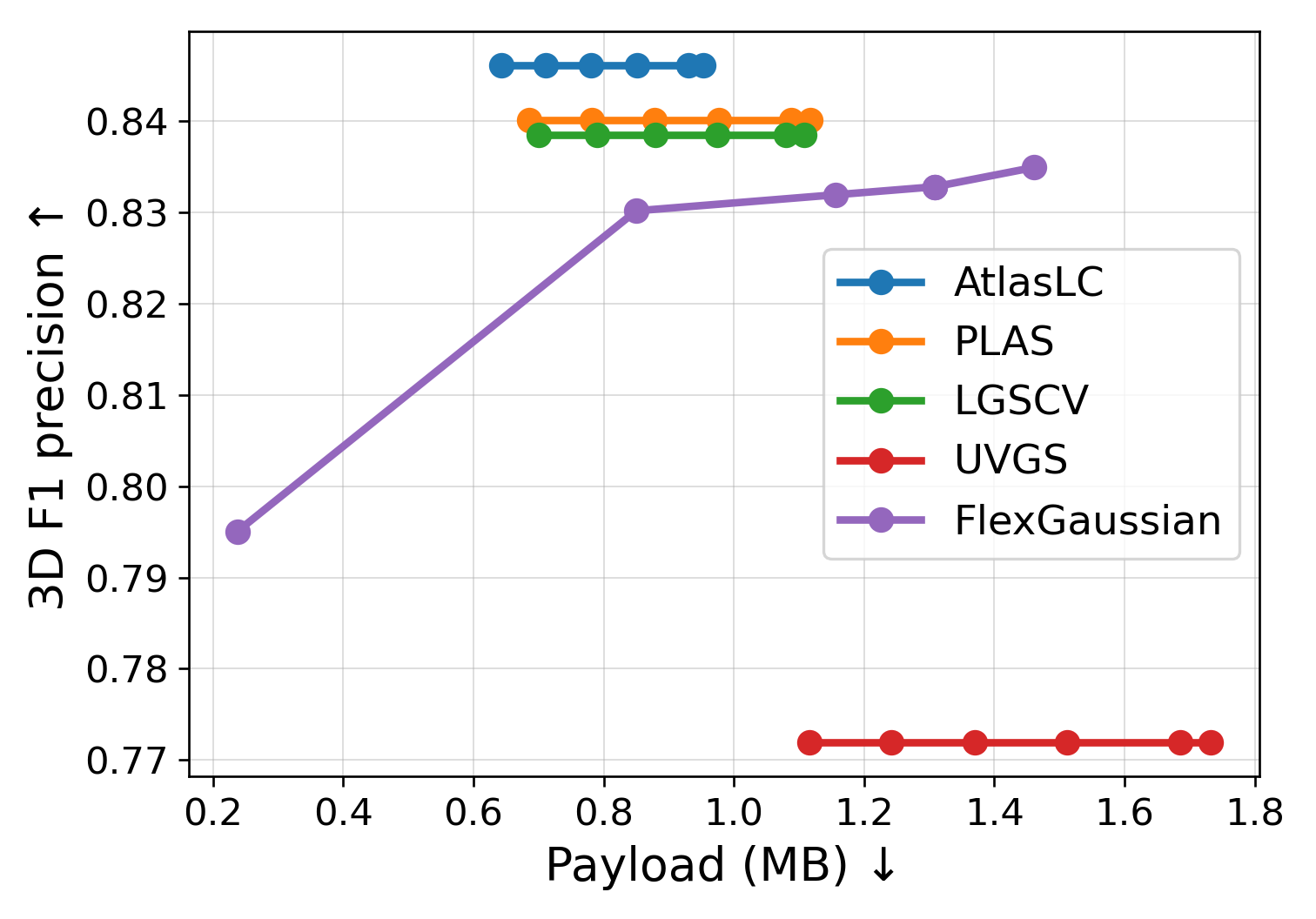}
    \vspace{-1.5mm}

    {\scriptsize \textbf{(c)} Payload--3D F1 precision.}
\end{minipage}\hfill
\begin{minipage}[t]{0.49\columnwidth}
    \centering
\includegraphics[
  width=\linewidth,
  alt={Panel d. Line chart of compressed payload versus three-dimensional F1 recall for AtlasLC, PLAS, LGSCV, UVGS, and FlexGaussian. Within the shared payload range, AtlasLC generally achieves higher recall than PLAS and LGSCV, indicating more complete coverage of the ground-truth object surface.}
]{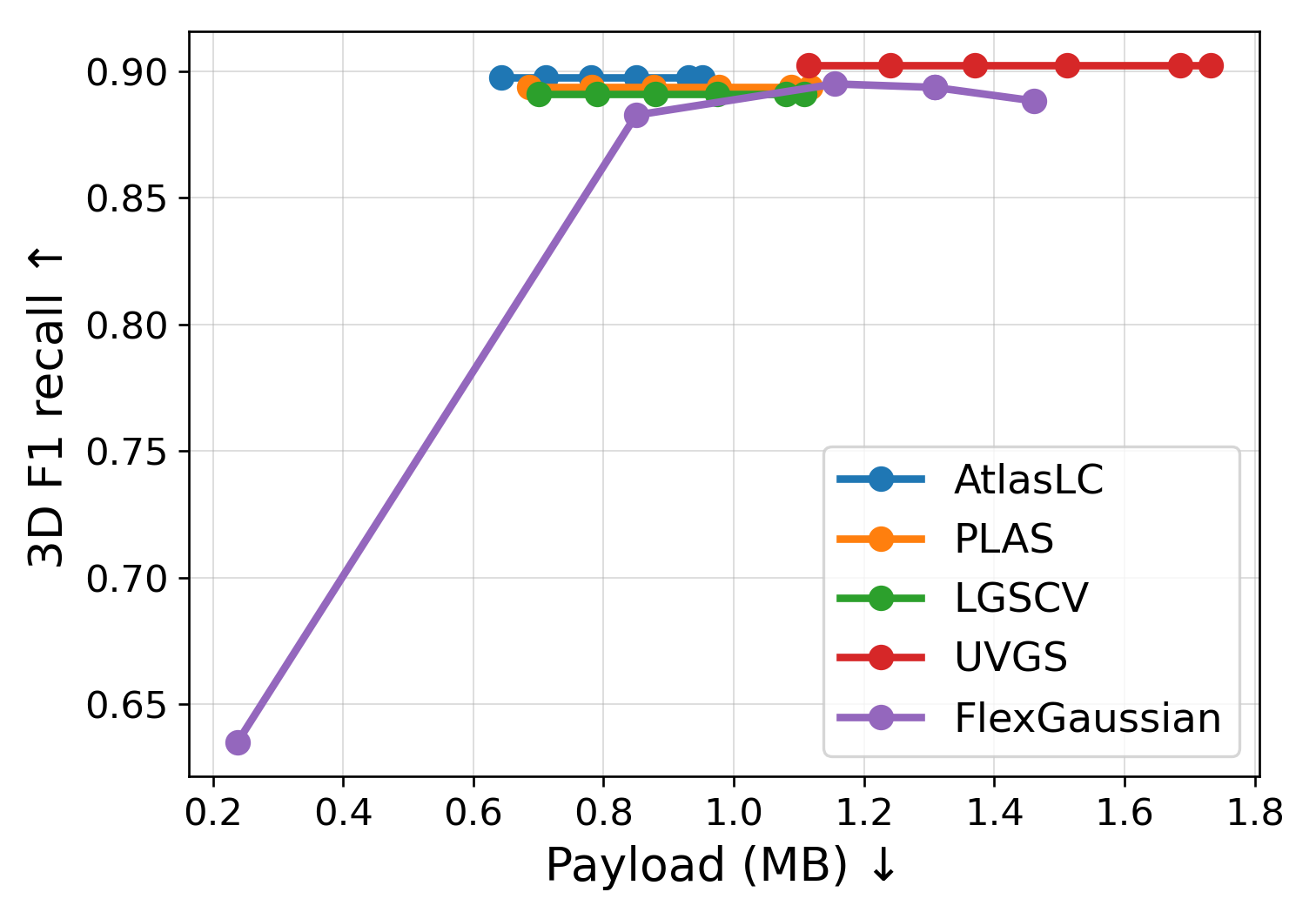}
    \vspace{-1.5mm}

    {\scriptsize \textbf{(d)} Payload--3D F1 recall.}
\end{minipage}
\vspace{-1mm}
\caption{
\textbf{Rate-dependent fidelity, geometry components, and deployment cost.}
We sweep HEVC QP $\{1,4,8,12,16,20\}$ for AtlasLC, PLAS,
LGSCV, and UVGS, and use the native rate ladder for FlexGaussian.
Payload--PSNR shows the image-space fidelity trade-off,
3D precision and recall decompose geometry preservation,
and payload--total time summarizes end-to-end deployment cost.
Within the common AtlasLC/PLAS/LGSCV payload range,
AtlasLC trades a small PSNR decrease for higher precision and
recall while maintaining substantially lower total pipeline time.
The timing curve is a deployment diagnostic rather than a strict
monotonic rate--distortion envelope because measured runtime can
vary across codec operating points.
}
\label{fig:rate_geometry}
\end{figure}

As shown in Fig.~\ref{fig:rate_geometry}, AtlasLC exhibits a
favorable rate-dependent trade-off on Objaverse.
Over the common AtlasLC/PLAS/LGSCV payload range, it incurs a
small PSNR decrease while improving both 3D precision and recall,
showing that the geometry gain is supported by both accurate
retained support and improved surface coverage.
The total-time curve further shows that AtlasLC maintains a
substantially lower end-to-end deployment cost across the
evaluated operating points.
Complementary stage-wise results in Sec.~\ref{sec:results} show
that AtlasLC reduces atlas-preparation time from roughly
900--1700\,ms for prior atlas-based pipelines to about 68\,ms,
yielding up to \textbf{25$\times$} faster atlas preparation and
\textbf{3--5$\times$} faster end-to-end compression.
Together, these results support deployment-aware object-centric
3DGS compression, where compactness, preparation cost, decoding
efficiency, and geometry preservation are considered jointly.

\textbf{Our contributions are threefold:}
\begin{itemize}
\item We identify object-centric, asset-only 3DGS compression as a distinct deployment setting in which mapping cost, decoding efficiency, and geometry preservation matter alongside bitrate.
\item We propose \textbf{AtlasLC}, which combines local-competition pruning with deterministic atlas packing over a lightweight shared ordering backbone, eliminating post-pruning remapping.
\item We evaluate compression under deployment-aware criteria spanning payload, total pipeline time, runtime efficiency, image fidelity, and geometry, showing that AtlasLC provides a favorable trade-off among the evaluated compressed baselines.
\end{itemize}

\section{Related Work}

\subsection{Post-Hoc 3DGS Compression and Compaction}

Since the introduction of 3D Gaussian Splatting (3DGS) for real-time radiance-field rendering \cite{Kerbl2023_3DGS}, a large body of work has studied how to reduce its storage and runtime cost. Recent surveys organize this space around pruning and compaction, quantization and entropy coding, and structured parameterization for codec compatibility \cite{Bagdasarian2025_3DGSzip,Yang2024_GGSCBenchmark}. Representative scene-scale compaction methods span pruning, vector quantization, entropy/context modeling, predictive coding, and rate--distortion optimization \cite{Papantonakis2024_ReducingMemoryFootprint3DGS,Fan2024_LightGaussian,Zhang2024_LP3DGS,Hanson2025_PUP3DGS,Wang2024_ContextGS,Chen2024_HAC,Liu2024_CompGS,Navaneet2024_CompGSVQ,Cao2025_LightweightPredictiveGS,Wang2024_RDOGaussian}. These methods demonstrate substantial bitrate reduction, but are primarily motivated by scene-scale storage reduction or rendering-oriented compaction.

A more deployment-relevant direction considers \emph{post-hoc} compression that operates on a trained Gaussian asset without the original capture workspace. MesonGS studies post-training compression through attribute transformation \cite{Xie2024_MesonGS}, FlexGaussian emphasizes training-free compression \cite{Tian2025_FlexGaussian}, and FCGS highlights fast compression-time behavior \cite{Chen2025_FCGS}. For reusable XR assets, this distinction is central: compression is often performed after asset release by a packaging or deployment stage without original images, camera poses, or scene-specific optimization. We build on this source-free, post-hoc viewpoint, but target a different bottleneck from most prior compaction methods: quickly turning an unordered object-centric Gaussian asset into a codec-ready representation for repeated deployment.

\subsection{Structured Layouts, Codec Compatibility, and Mapping Cost}

A complementary line of work imposes spatial structure on unordered Gaussians so that standard image or video codecs can exploit local correlation. Compressed 3D Gaussian Splatting showed that compressed splat representations can reduce bandwidth and rendering cost for accelerated novel-view synthesis \cite{Niedermayr2024_Compressed3DGS}. More recent methods organize Gaussian attributes into 2D layouts or codec-friendly parameterizations: Self-Organizing Gaussian Grids use PLAS to arrange Gaussians
into coherent grids
\cite{Morgenstern2024_SelfOrganizingGaussianGrids}.
UVGS maps unordered Gaussians to a regular UV domain using
spherical projection~\cite{Rai2025_UVGS}, while OT-UVGS
reframes UV assignment as a fixed-capacity allocation problem
and applies a lightweight rank-based separable OT-inspired
mapping to improve UV-slot utilization and reduce collisions
without changing the underlying UVGS representation
\cite{Kim2026_OTUVGS}.
CodecGS uses optimized feature planes for standard video codecs
\cite{Lee2025_CodecGS}, LGSCV and VQ/HEVC-style pipelines
pursue lighter map generation or hybrid coding
\cite{Yang2025_LGSCV,Kim2025_VQHEVCHybrid3DGS}, and recent
work studies global size allocation or hierarchical atlas
optimization for rate--distortion improvement
\cite{Xie2025_SizeGS,Wang2025_GHAP3DGS}.

These methods show that codec-friendly layouts improve compactness and decoding practicality, but the layout stage itself can become a preprocessing bottleneck when it requires iterative self-organization, UV assignment, feature-plane optimization, mixed-integer size allocation, or extra remapping \cite{Morgenstern2024_SelfOrganizingGaussianGrids,Rai2025_UVGS,Lee2025_CodecGS,Yang2025_LGSCV,Xie2025_SizeGS,Wang2025_GHAP3DGS}. This cost matters for object-centric XR asset pipelines, where many reusable objects must be packaged, refreshed, and redeployed repeatedly rather than amortized over a few offline scenes. AtlasLC is motivated by this gap: it keeps the practical advantage of codec-compatible atlases, but replaces heavier mapping/remapping with local-competition pruning and deterministic atlas packing over a lightweight shared ordering.

\subsection{Object-Centric 3DGS Assets and Deployment-Aware Evaluation}

Beyond general 3DGS compression, radiance fields are increasingly treated as deployable XR content rather than only offline scene reconstructions \cite{Li2025_RadianceFieldsXR}. This shift is reflected in large object repositories such as Objaverse \cite{Deitke2023_Objaverse}, object-centric datasets and benchmarks such as DTC \cite{Dong2025_DTC} and uCO3D \cite{Liu2025_uCO3D}, and generative asset pipelines for reusable 3D content \cite{Yang2025_GenAssets}. At the scene-composition level, recent spatial-MR authoring
systems generate layout-consistent 3D scenes from RGB sequences
through object-level semantic scene graphs, enabling virtual
content to reflect user-specific physical environments
\cite{Kim2026_SceneLinker}. At the systems level, scene delivery, adaptive streaming, XR traffic, and remote rendering have also become first-class concerns for volumetric and Gaussian content \cite{Liu2022_Vues,Tsai2025_L3GS,Shi2025_LapisGS,Alnajim2023_XRTraffic,Gao2025_XRgo,Luo2023_3GPPXR}. In this setting, the practical unit is often a reusable object asset that must be prepared, transmitted, decoded, cached, and instantiated quickly across heterogeneous XR systems \cite{Dogan2024_XRObjects,Friston2017_3DRepo4Unity}.

This operating regime changes how compression should be designed and evaluated. Compared with scene-scale captures, object-centric assets devote most of their Gaussian budget to a single foreground instance, so removable redundancy is often \emph{local rather than globally disposable}. Scene-scale pruning methods such as LightGaussian, LP-3DGS, PUP, Fragment Pruning, and GoDe are effective for compaction \cite{Fan2024_LightGaussian,Zhang2024_LP3DGS,Hanson2025_PUP3DGS,Ye2024_FragmentPruning,DiSario2025_GoDe}, but object assets place more pressure on preserving silhouettes, thin structures, and depth-consistent support. Geometry-aware Gaussian representations such as 2DGS, SuGaR, Gaussian Frosting, GaussianUDF, MaGS, and ArtGS likewise show that depth fidelity and surface consistency matter beyond RGB reconstruction \cite{Huang2024_2DGS,Guedon2024_SuGaR,Guedon2024_GaussianFrosting,Li2025_GaussianUDF,Ma2025_MaGS,Liu2025_ArtGS}.

XR studies on occlusion, positioning, and interaction further motivate preserving these geometry cues \cite{Macedo2023_OcclusionHandlingAR,Walton2017_AccurateOcclusionMR,Hellmuth2021_MobileARPositioning,Lilija2019_OccludedInteraction,Tang2020_GrabAR}. We therefore evaluate object-centric 3DGS compression not only by payload and image fidelity, but also by total pipeline time, runtime efficiency, and geometry preservation. While structured object-centric methods such as UVGS point in this direction \cite{Rai2025_UVGS}, the combination of source-free compression, low mapping cost, and deployment-aware evaluation remains underexplored.

\begin{table}[t]
\centering
\caption{
\textbf{Baseline assumptions relative to AtlasLC.}
\cmark: native support; \pmark: partial or adapted support; \xmark: not native; --: not applicable because the method does not construct an atlas/UV map.
``Src.'' denotes strict source-free compression without images, poses, rendered validation views, or per-object quality feedback.
PUP is included because it is used only in the prune-only diagnostic.
}
\label{tab:baseline_assumptions_compact}
\scriptsize
\setlength{\tabcolsep}{2.1pt}
\renewcommand{\arraystretch}{0.96}
\resizebox{\columnwidth}{!}{%
\begin{tabular}{lcccccc}
\toprule
\textbf{Method}
& \textbf{Obj.}
& \textbf{Src.}
& \textbf{No train}
& \textbf{Std. codec}
& \textbf{Atlas/UV}
& \textbf{No opt./remap} \\
\midrule
PLAS/SOG~\cite{Morgenstern2024_SelfOrganizingGaussianGrids}
& \xmark & \pmark & \pmark & \cmark & \cmark & \xmark \\
LGSCV~\cite{Yang2025_LGSCV}
& \xmark & \cmark & \cmark & \cmark & \cmark & \pmark \\
UVGS~\cite{Rai2025_UVGS}
& \cmark & \pmark & \pmark & \pmark & \cmark & \pmark \\
FlexGaussian~\cite{Tian2025_FlexGaussian}
& \xmark & \cmark & \cmark & \xmark & \xmark & -- \\
PUP~\cite{Hanson2025_PUP3DGS}
& \xmark & \xmark & \xmark & \xmark & \xmark & -- \\
\midrule
\textbf{AtlasLC}
& \cmark & \cmark & \cmark & \cmark & \cmark & \cmark \\
\bottomrule
\end{tabular}%
}
\vspace{-2mm}
\end{table}

Table~\ref{tab:baseline_assumptions_compact} positions the baselines used in our experiments.
PLAS and LGSCV are structured standard-codec layout baselines, UVGS is an object-structured UV-domain baseline, FlexGaussian is a training-free parameter-compression baseline, and PUP is used only for the prune-only diagnostic.
AtlasLC targets the combined constraint of source-free, training-free, standard-codec-compatible atlas construction for reusable object-centric XR assets.

\section{Method}

We present \textbf{AtlasLC}, a training-free compression pipeline for object-centric 3D Gaussian Splatting.
The method is designed for the deployment regime: reusable object assets must be compressed \emph{post hoc}, without source data or per-asset optimization, and prepared quickly enough for repeated packaging and redeployment.
AtlasLC therefore focuses on two deployment-critical operations:
\emph{local-competition pruning}, which removes redundancy while preserving object-wide foreground support, and
\emph{deterministic atlas packing}, which converts the retained Gaussians into a stable, codec-friendly representation with low mapping overhead.
A lightweight sort-based conditional transport is used as a shared coordinate backbone for these stages; it is not an optimization-heavy mapping objective in itself.

\subsection{Problem Setup}

Let
\[
G=\{g_i\}_{i=1}^{N}
\]
denote an object-centric 3DGS model.
Each Gaussian \(g_i\) has a 3D center \(\mathbf{x}_i\in\mathbb{R}^3\), opacity \(\alpha_i\in\mathbb{R}_{+}\), and an attribute vector \(\mathbf{p}_i\in\mathbb{R}^{C}\) containing the parameters to be encoded, such as scale, rotation, color, and higher-order appearance coefficients.

Our goal is to transform the unordered set \(G\) into a codec-compatible atlas of spatial resolution \(H\times W\) with \(K\) slots per pixel.
We represent the final discrete assignment by an index tensor
\[
M\in\{0,1,\dots,N\}^{H\times W\times K},
\]
where \(M(r,c,k)=i\) indicates that slot \(k\) at pixel \((r,c)\) stores Gaussian \(g_i\), and \(0\) denotes an empty token.
The total atlas capacity is
\[
T = HWK.
\]

AtlasLC therefore seeks
(i) a retained index set
\[
S\subseteq \{1,\dots,N\},
\]
and
(ii) a deterministic, collision-free mapping from the retained indices in \(S\) into \(M\),
such that the resulting atlas is compact, codec-friendly, and conservative with respect to the foreground support needed for appearance, silhouettes, and stable depth.
Under our deployment setting, the method should also keep asset-preparation cost low:
the atlas should be constructed without iterative remapping, scene-specific learning, or expensive global assignment.

\subsection{Overview}

AtlasLC consists of three stages:
\begin{enumerate}
    \item \textbf{Lightweight shared ordering.} We build a cheap sort-based conditional 2D ordering over the input Gaussians. Its role is to regularize the unordered set into a stable coordinate backbone that can be reused downstream.
    \item \textbf{Local-competition pruning.} We quantize this ordering into coarse competition cells and prune \emph{within} each cell using local quotas, so that redundancy is removed locally rather than by a single global top-\(k\) rule.
    \item \textbf{Deterministic atlas packing.} The retained Gaussians are packed directly into the atlas using a stable, collision-free packing policy that preserves locality and avoids additional remapping.
\end{enumerate}

AtlasLC is faster than prior atlas-based pipelines because it replaces iterative layout construction, post-pruning remapping, and optimization-heavy refinement with a \emph{single shared ordering} and \emph{direct deterministic packing}. The ordering is computed once and reused by both pruning and packing, so atlas preparation does not require repeated self-organization or a second map-generation stage. At the same time, AtlasLC remains competitive in quality because the shared ordering still regularizes global density for codec efficiency, local-competition pruning removes redundancy only within coarse neighborhoods rather than globally eroding foreground support, and deterministic packing preserves local attribute smoothness that standard codecs can exploit.

\subsection{Lightweight Shared Ordering Backbone}

\subsubsection{Base parameterization}

For each Gaussian \(g_i\), we first compute a deterministic base coordinate
\[
\mathbf{u}_i^0=(u_i^0,v_i^0)\in [0,1)^2.
\]
In all experiments, we use a deterministic spherical parameterization. The method only requires the initializer to be deterministic and inexpensive.

\subsubsection{Sort-based Conditional rank transform}

We convert the base coordinates into structured 2D coordinates
\[
\mathbf{u}_i=(u_i,v_i)\in[0,1)^2
\]
using weighted rank transforms with weights
\[
w_i=\max(\alpha_i,\varepsilon)^\gamma.
\]
After choosing a seam shift \(\delta^\star\) that minimizes weighted mass near the wrap boundary, we apply a global weighted CDF along \(u\) and a conditional weighted CDF along \(v\) within coarse \(u\)-bins:
\[
u_i = F_U\!\left((u_i^0-\delta^\star)\bmod 1\right),
\qquad
v_i = F_{V|U}\!\left(v_i^0 \mid \mathrm{bin}(u_i)\right).
\]
For a sorted one-dimensional sequence, we use the weighted mid-rank transform
\[
F(x_{(t)};w)=
\frac{\sum_{\ell=1}^{t} w_{(\ell)}-\frac{1}{2}w_{(t)}}{\sum_{\ell=1}^{n} w_{(\ell)}}.
\]
This step is used only to produce a stable shared ordering; it is not itself an optimization-heavy transport objective.

\begin{figure*}[t]
\centering
\includegraphics[
  width=\textwidth,
  alt={Side-by-side visualization of Gaussians removed from the same object-centric 3D Gaussian Splatting asset under matched pruning budgets. The left view marks Gaussians pruned by PUP in red, distributed broadly across visible object-surface support. The right view marks Gaussians pruned by AtlasLC in yellow, concentrated in compact local groups. The comparison shows that AtlasLC removes local redundancy while preserving broader object coverage and geometry-critical support.}
]{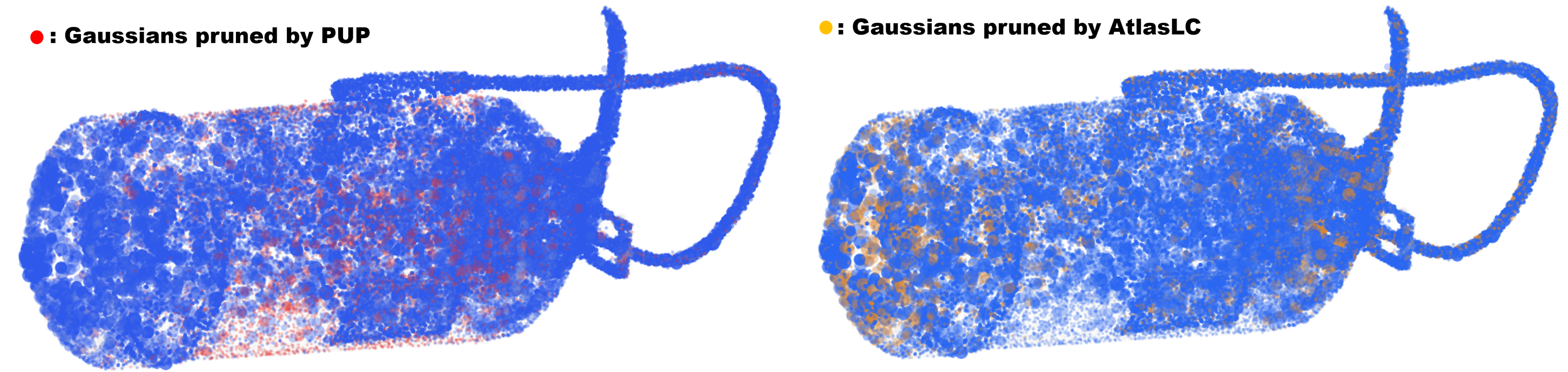}\caption{
Visualization of prune-only behavior on an object-centric 3DGS asset.
\textbf{Red} points denote Gaussians pruned by PUP~\cite{Hanson2025_PUP3DGS} (left), \textbf{yellow} points denote Gaussians pruned by AtlasLC local-competition pruning (right).
PUP removes splats more diffusely across the object support, including points spread over visible surface regions.
By contrast, AtlasLC removes Gaussians in more spatially concentrated local groups, indicating that it suppresses \emph{locally redundant} splats while preserving broader surface coverage and geometry-critical support.
This behavior is consistent with our local-competition design and helps explain why AtlasLC better preserves object surface structure and geometry under matched pruning budgets.
}
\label{fig:pup_vs_scotlc_pruning}
\end{figure*}

\begin{table}[t]
\centering
\small
\setlength{\tabcolsep}{5.0pt}
\renewcommand{\arraystretch}{1.08}
\caption{
Prune-only comparison between \textbf{PUP}~\cite{Hanson2025_PUP3DGS} and \textbf{AtlasLC local-competition pruning} for pruning ratios from 30\% to 90\%.
\( |\Delta G| \) denotes the absolute difference in kept-Gaussian counts between the two methods.
Metric values are $\Delta$(\textbf{AtlasLC} $-$ \textbf{PUP}); positive favors AtlasLC for PSNR/SSIM/3D F1, and negative favors AtlasLC for LPIPS.
}
\label{tab:pup_atlaslc_prune_only_30_90_compact}
\begin{tabular}{c c c c c c}
\toprule
\textbf{Prune}
& \textbf{$|\Delta G|$}
& \textbf{$\Delta$PSNR}$\uparrow$
& \textbf{$\Delta$SSIM}$\uparrow$
& \textbf{$\Delta$LPIPS}$\downarrow$
& \textbf{$\Delta$3D F1}$\uparrow$ \\
\midrule
90\% & 0 & +0.498 & +0.068 & -0.015 & +0.053 \\
80\% & 0 & +0.592 & +0.061 & -0.013 & +0.063 \\
70\% & 1 & +0.434 & +0.041 & -0.010 & +0.024 \\
60\% & 1 & +0.311 & +0.015 & -0.005 & +0.008 \\
50\% & 0 & +0.298 & +0.005 & -0.001 & +0.010 \\
40\% & 0 & +0.018 & +0.001 & +0.001 & +0.029 \\
30\% & 0 & +0.015 & +0.001 & +0.002 & +0.041 \\
\bottomrule
\end{tabular}
\vspace{-1mm}
\end{table}

\subsection{Local-Competition Pruning}

\subsubsection{Competition cells}

We quantize the shared ordering into a coarse pruning grid of \(B_x^{p}\times B_y^{p}\) cells:
\[
c_i=
\left(
\left\lfloor B_x^{p}u_i\right\rfloor,
\left\lfloor B_y^{p}v_i\right\rfloor
\right).
\]
Each cell defines a \emph{local competition neighborhood}.
Gaussians assigned to the same cell are interpreted as likely competitors for nearby atlas support.

This local view is especially important for object-centric assets.
A global pruning rule tends to over-allocate budget to already dense regions and can erase thin structures or boundary support elsewhere on the object.
By contrast, cell-wise competition removes redundancy \emph{inside} neighborhoods while preserving support diversity \emph{across} the object.

\subsubsection{Input-adaptive budget and quota allocation}
\label{sec:adaptive_budget}

For an asset \(a\) with \(N_a\) input Gaussians, AtlasLC computes the keep ratio and target survivor count as
\[
\begin{aligned}
\rho_a
&=
\min\left(
\operatorname{clip}\left(\frac{M_0}{N_a},\rho_{\min},\rho_{\max}\right),
\frac{\beta T}{N_a}
\right),\\
m_a
&=
\min\left(\left\lfloor \rho_a N_a \right\rceil,T\right).
\end{aligned}
\]
Here, \(M_0\), \(\rho_{\min}\), \(\rho_{\max}\), and \(\beta\) are fixed globally for the codec-atlas configuration. The budget depends only on \(N_a\) and the atlas capacity \(T=HWK\); it uses no source images, camera poses, rendered views, PSNR, SSIM, LPIPS, 3D F1, depth, or per-object validation feedback. Thus, ``adaptive'' denotes input adaptation rather than per-object quality tuning.

Let \(n_{a,c}\) denote the occupancy of competition cell \(c\) for asset \(a\). Its ideal quota is
\[
\hat q_{a,c}=\rho_a n_{a,c}.
\]
We convert the ideal quotas into bounded integer quotas \(q_{a,c}\) using largest-remainder apportionment, subject to
\[
0\le q_{a,c}\le n_{a,c},
\qquad
\sum_c q_{a,c}\approx m_a.
\]
When feasible, one survivor is assigned to every non-empty cell so that isolated support is not removed entirely. This distributes survivor capacity across the object rather than allowing a single global ranking to absorb the budget into the densest regions.

\subsubsection{Local survivor selection}

Within each cell, Gaussians are ranked by a lightweight value score \(s_i\).
In all experiments, we use the fixed opacity score \(s_i=\alpha_i\), without per-object tuning or validation feedback.
The retained set is then selected as
\[
S
=
\bigcup_c
\operatorname{TopK}
\Big(
\{i \mid c_i=c\},
q_{a,c};\; s_i
\Big).
\]

If rounding produces a slight budget mismatch, we resolve it globally by trimming or filling according to the same score \(s_i\).
Finally, we enforce the atlas-capacity constraint
\[
|S| \le T.
\]

This local-competition formulation is one of the two core components of AtlasLC.
It does not ask which Gaussians are globally most salient in isolation; instead, it asks which Gaussians are most valuable \emph{relative to their local competitors}.
That bias is particularly appropriate for object assets, where many splats are locally redundant but globally important for preserving a semantically concentrated foreground object.

\begin{table*}[t]
\centering
\caption{Main comparison on the 320-object Objaverse evaluation set. Best values are in \textbf{bold}; second-best values are \underline{underlined}. N/A indicates that the corresponding stage is not directly comparable; for such rows, Total sums only the comparable stages. FLEX* denotes FlexGaussian~\cite{Tian2025_FlexGaussian}.}
\label{tab:main_comparison}
\small
\setlength{\tabcolsep}{2.5pt}
\resizebox{\textwidth}{!}{%
\begin{tabular}{lccccccccccc}
\toprule
Method & Payload (MB)$\downarrow$ & PSNR (dB)$\uparrow$ & SSIM$\uparrow$ & LPIPS$\downarrow$ & 3D F1$\uparrow$ & Depth RMSE$\downarrow$ & Map (ms)$\downarrow$ & Encode (ms)$\downarrow$ & Decode (ms)$\downarrow$ & FPS$\uparrow$ & Total (ms)$\downarrow$ \\
\midrule
VANILLA~\cite{Kerbl2023_3DGS} & 22.223 & \textbf{28.078} & \textbf{0.9461} & \textbf{0.0579} & \textbf{0.870} & \textbf{0.0310} & N/A & N/A & N/A & 129.8 & N/A \\
PLAS~\cite{Morgenstern2024_SelfOrganizingGaussianGrids}       & \underline{0.686} & \underline{27.226} & \underline{0.9406} & 0.0740 & 0.860 & \underline{0.0350} & 1731.0 & \textbf{170.8} & \underline{122.7} & 130.2 & 2024.4 \\
LGSCV~\cite{Yang2025_LGSCV}        & 0.700 & 27.132 & 0.9394 & 0.0744 & 0.858 & 0.0354 & 911.7 & 194.6 & 127.0 & 131.7 & 1233.3 \\
UVGS~\cite{Rai2025_UVGS}         & 1.370 & 24.373 & 0.8857 & 0.0922 & 0.827 & 0.0514 & \textbf{43.8} & 216.7 & 133.2 & 128.3 & \underline{393.7} \\
FLEX*~\cite{Tian2025_FlexGaussian}        & 0.850 & 25.739 & 0.9026 & 0.0749 & 0.851 & 0.0497 & N/A & 334.5 & 141.4 & \underline{131.9} & 475.9 \\
\midrule
\textbf{AtlasLC} & \textbf{0.643} & 27.034 & 0.9301 & \underline{0.0732} & \underline{0.865} & 0.0361 & \underline{68.5} & \underline{189.0} & \textbf{108.1} & \textbf{135.8} & \textbf{365.6} \\
\bottomrule
\end{tabular}%
}
\end{table*}

\begin{table*}[t]
\centering
\caption{Evaluation on the 30-object Stanford-ORB dataset. Best values are in \textbf{bold}; second-best values are \underline{underlined}. N/A indicates that the corresponding stage is not directly comparable; for such rows, Total sums only the comparable stages. FLEX* denotes FlexGaussian~\cite{Tian2025_FlexGaussian}.}
\label{tab:main_comparison_stanfordorb}
\small
\setlength{\tabcolsep}{2.5pt}
\resizebox{\textwidth}{!}{%
\begin{tabular}{lccccccccccc}
\toprule
Method & Payload (MB)$\downarrow$ & PSNR (dB)$\uparrow$ & SSIM$\uparrow$ & LPIPS$\downarrow$ & 3D F1$\uparrow$ & Depth RMSE$\downarrow$ & Map (ms)$\downarrow$ & Encode (ms)$\downarrow$ & Decode (ms)$\downarrow$ & FPS$\uparrow$ & Total (ms)$\downarrow$ \\
\midrule
VANILLA~\cite{Kerbl2023_3DGS} & 28.014 & \textbf{30.808} & \textbf{0.9573} & \textbf{0.0147} & \textbf{0.650} & \textbf{0.0604} & N/A & N/A & N/A & 44.4 & N/A \\
PLAS~\cite{Morgenstern2024_SelfOrganizingGaussianGrids} & 1.376 & 27.276 & 0.8964 & 0.0244 & 0.617 & \underline{0.0848} & 1575.3 & 297.6 & 222.3 & 44.8 & 2095.3 \\
LGSCV~\cite{Yang2025_LGSCV} & \underline{1.273} & 27.253 & 0.8963 & 0.0245 & 0.618 & 0.0908 & 1069.8 & \underline{292.0} & \underline{221.7} & 45.5 & 1583.5 \\
UVGS~\cite{Rai2025_UVGS} & 2.516 & 24.099 & 0.8343 & 0.0308 & 0.621 & 0.2337 & \textbf{59.0} & 329.1 & 247.8 & \underline{53.4} & \underline{635.8} \\
FLEX*~\cite{Tian2025_FlexGaussian} & 1.616 & \underline{27.523} & \underline{0.9027} & \underline{0.0225} & 0.634 & 0.1249 & N/A & 484.6 & 232.6 & 46.8 & 717.1 \\
\midrule
\textbf{AtlasLC} & \textbf{1.080} & 27.016 & 0.8814 & 0.0277 & \underline{0.647} & 0.1001 & \underline{97.4} & \textbf{279.2} & \textbf{207.5} & \textbf{58.4} & \textbf{584.2} \\
\bottomrule
\end{tabular}%
}
\end{table*}

\begin{table*}[t]
\centering
\caption{Ablation of the two main design choices, with the shared ordering backbone kept fixed. Deterministic packing is essential for compact, codec-efficient deployment, while local-competition pruning further improves compactness and geometry under the same packing backbone. Best values are shown in \textbf{bold}; second-best values are \underline{underlined}.}
\label{tab:ablation_scot_lc_det_no_decode_outliers}
\small
\setlength{\tabcolsep}{2.6pt}
\resizebox{\textwidth}{!}{%
\begin{tabular}{lcc|ccccccccccccc}
\toprule
Variant 
& LC prune
& Det. pack 
& Payload (MB)$\downarrow$
& PSNR (dB)$\uparrow$
& SSIM$\uparrow$
& LPIPS$\downarrow$
& 3D F1$\uparrow$
& Depth RMSE$\downarrow$
& Map (ms)$\downarrow$
& Encode (ms)$\downarrow$
& Decode (ms)$\downarrow$
& FPS$\uparrow$
& Total (ms)$\downarrow$ \\
\midrule
Deterministic packing only
& \xmark & \cmark
& \underline{1.003}
& \textbf{27.314}
& \textbf{0.9393}
& \textbf{0.0725}
& \underline{0.864}
& \underline{0.0366}
& \textbf{51.7}
& \underline{197.9}
& \underline{118.2}
& 129.3
& \underline{367.8} \\

LC pruning only
& \cmark & \xmark
& 2.848
& 25.195
& 0.8820
& 0.0902
& 0.853
& 0.0382
& 76.6
& 224.8
& 152.7
& \underline{134.1}
& 454.1 \\

AtlasLC
& \cmark & \cmark
&\textbf{0.643} & \underline{27.034} & \underline{0.9301} & \underline{0.0732} & \textbf{0.865} & \textbf{0.0361} & \underline{68.5} & \textbf{189.0} & \textbf{108.1} & \textbf{135.8} & \textbf{365.6} \\
\bottomrule
\end{tabular}%
}
\end{table*}

\subsection{Deterministic Atlas Packing}

After pruning, AtlasLC performs \emph{deterministic atlas packing} over the retained coordinates
\[
\{\mathbf{u}_i \mid i\in S\}.
\]
This is the deployment-critical stage that turns the retained Gaussian set into a practical codec asset.
Because the retained set is already capped by atlas capacity, the role of packing is not to decide what to keep, but to place the retained Gaussians into the atlas in a stable, collision-free, and codec-efficient way.

\subsubsection{Coarse packing bins}

We quantize the retained coordinates into \(B_x\times B_y\) coarse packing bins:
\[
b_i=
\left(
\left\lfloor B_x u_i\right\rfloor,
\left\lfloor B_y v_i\right\rfloor
\right),
\qquad i\in S.
\]
Each bin \(b\) corresponds to a rectangular atlas region \(R_b\).
The slot capacity of that region is
\[
C_b = |R_b|K.
\]

This coarse binning step converts the shared ordering into a discrete packing prior.
It preserves broad spatial regularity without requiring any iterative reassignment or optimization-heavy layout refinement.

\subsubsection{Stable point and slot orders}

Let
\[
S_b=\{i\in S\mid b_i=b\}
\]
be the set of retained Gaussians assigned to block \(b\).
AtlasLC uses two deterministic orders:
\begin{itemize}
    \item a \emph{slot order} over the pixels and \(K\) slots within \(R_b\), and
    \item a \emph{point order} over the Gaussians in \(S_b\).
\end{itemize}

In all experiments, the slot order is raster scan and the point order is 3D Morton order.
These choices keep the packing stage cheap while preserving local attribute smoothness that standard codecs can exploit.

Unlike heavier mapping pipelines, AtlasLC does not perform a second learned or optimization-based remapping after pruning.
The retained Gaussians are placed directly according to these stable orders, which makes the atlas deterministic across runs and keeps map generation cost low.

\subsubsection{Overflow handling}

If a block is overfull, i.e., \(|S_b| > C_b\), the excess Gaussians form an overflow set \(O_b\).
We concatenate all overflow sets into a global overflow set
\[
O = \bigcup_b O_b,
\]
and then place these remaining Gaussians into empty atlas slots in a deterministic global order.

This two-stage procedure yields a collision-free mapping by construction.
The shared ordering regularizes the retained distribution before placement, while deterministic packing resolves the discrete capacity constraints of the atlas with negligible additional overhead.
Across the evaluated assets, 2.55\% of the \(16\times16\) packing bins were temporarily overfull before spill handling, and 1.95\% of retained Gaussians were temporarily relocated. After deterministic global spill placement, the final collision, dropped-retained-Gaussian, and unplaced-retained-Gaussian counts were all zero. Temporary overflow is therefore a local placement event rather than retained-Gaussian loss.
The final output is an index map \(M\) together with the gathered payload atlas, which can be quantized and encoded directly with a standard image or video codec.

\subsection{Complexity and Practical Properties}

AtlasLC is entirely training-free.
It uses only geometric initialization, weighted sorting, cell histogramming, quota allocation, and deterministic packing.
No learned prior, no scene-specific adaptation, and no Sinkhorn-style transport optimization are required.

Let \(N\) be the number of input Gaussians.
The dominant cost arises in shared-ordering construction, where weighted sorting determines the overall time complexity of the pipeline as
\[
O(N\log N).
\]
The subsequent stages do not change this bound:
cell formation and quota allocation are linear in the number of Gaussians,
local survivor selection is linear once elements are organized by cell and score,
and deterministic packing is linear in the number of retained Gaussians \(|S|\) and the atlas size.
Thus, after the initial sorting-dominated stage, the rest of the pipeline remains lightweight in practice.

Two practical properties are especially important for XR deployment.
First, AtlasLC is \emph{codec-compatible by construction}: the output is an ordered atlas representation that can be directly encoded by off-the-shelf image or video codecs.
Second, AtlasLC is \emph{deployment-oriented by design}: local-competition pruning preserves object-wide foreground support, while deterministic packing removes the heavy mapping/remapping bottleneck that otherwise inflates asset-preparation and total compression pipeline time.

\begin{figure*}[t]
\centering
\includegraphics[
  width=\textwidth,
  alt={Grid of RGB renderings for three representative Objaverse objects. Each object is compared using the uncompressed Vanilla representation, PLAS, AtlasLC, UVGS, and FlexGaussian at representative operating points. AtlasLC remains visually close to Vanilla and PLAS, whereas UVGS and FlexGaussian exhibit more visible boundary degradation on thin structures, including the narrow ankle region of the boot.}
]{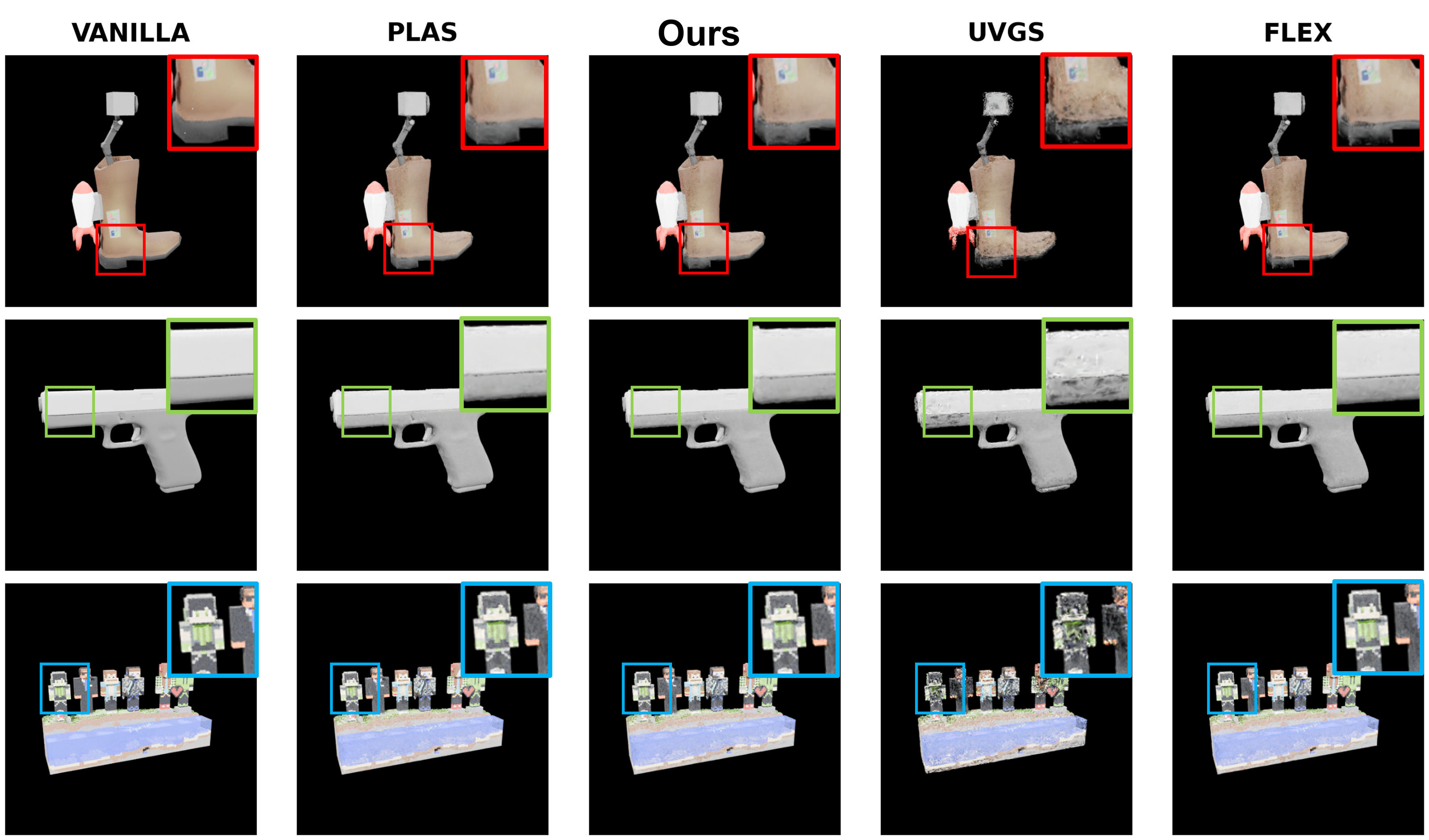}\caption{
\textbf{Representative RGB comparison on three Objaverse assets.}
We compare \textbf{VANILLA}, \textbf{PLAS}, \textbf{AtlasLC (Ours)}, \textbf{UVGS}, and \textbf{FLEX} using RGB renderings.
AtlasLC remains visually close to VANILLA and PLAS, while UVGS and FLEX* show slight boundary degradation in thin structures (e.g., the narrow ankle region of the boot).
Corresponding depth-error visualizations are provided in Figure~\ref{fig:depth_compare}.
}
\label{fig:rgb_compare}
\end{figure*}

\begin{figure*}[t]
\centering
\includegraphics[
  width=\textwidth,
  alt={Grid of per-pixel depth-error visualizations for three representative Objaverse objects. Each object is compared using the uncompressed Vanilla representation, PLAS, AtlasLC, UVGS, and FlexGaussian at representative operating points. AtlasLC shows cleaner object boundaries and lower depth error than UVGS and FlexGaussian, particularly around thin and detailed structures.}
]{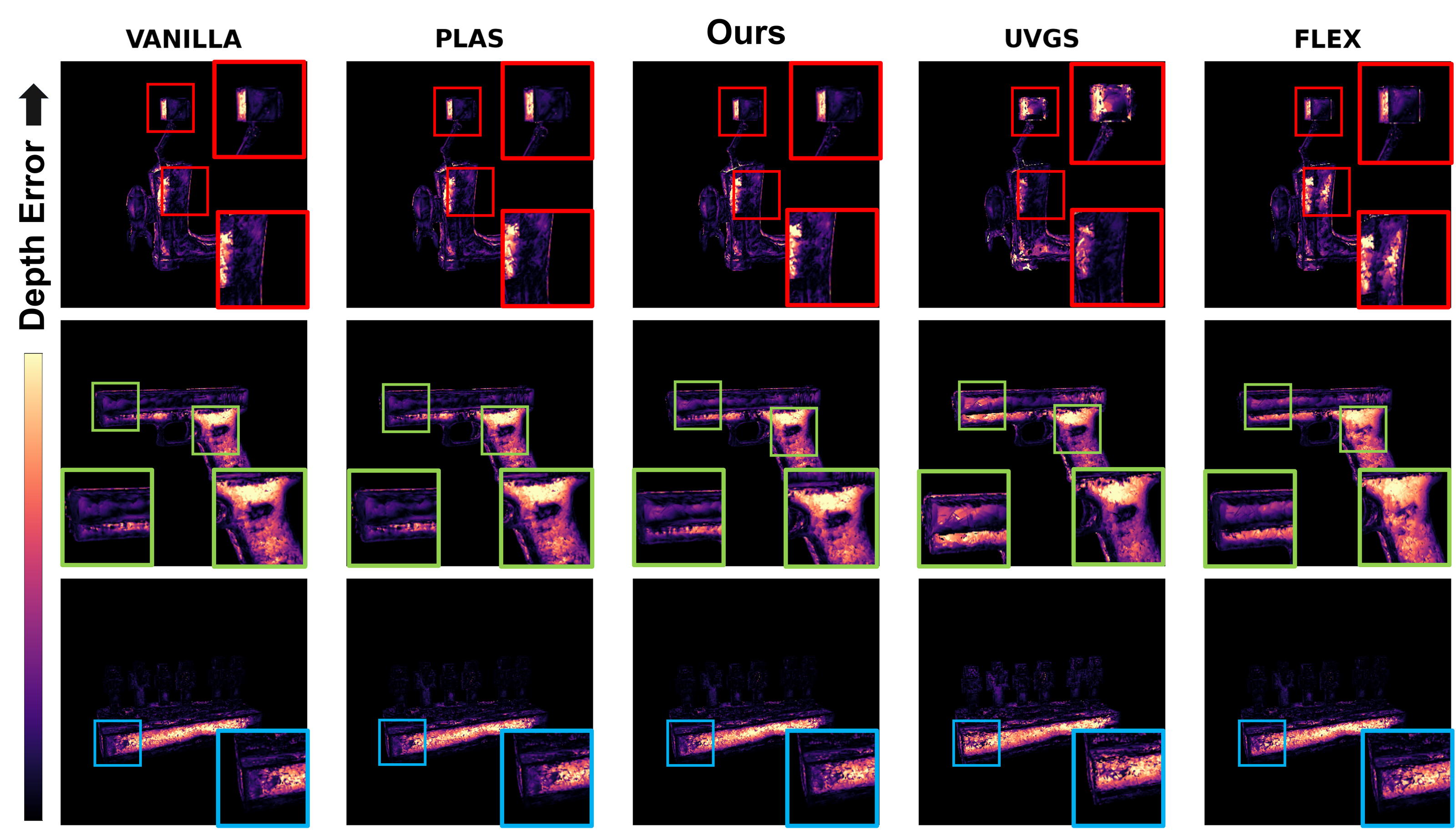}\caption{
\textbf{Representative depth-error comparison on three Objaverse assets.}
We compare \textbf{VANILLA}, \textbf{PLAS}, \textbf{AtlasLC (Ours)}, \textbf{UVGS}, and \textbf{FLEX} using per-pixel depth-error visualizations.
AtlasLC exhibits cleaner depth boundaries and lower depth error than UVGS and FLEX, particularly around thin and detailed structures.
Methods are shown at the representative operating points used in Table~\ref{tab:main_comparison}.
}
\label{fig:depth_compare}
\end{figure*}

\section{Results}
\label{sec:results}

We evaluate \textbf{AtlasLC} along three complementary axes.
We first report the full end-to-end comparison because AtlasLC is proposed as a \emph{deployment-oriented pipeline}, not as a pruning-only method.
We then isolate the contribution of \emph{local-competition pruning} under matched survivor budgets, and finally ablate the two main design choices---\emph{deterministic atlas packing} and \emph{local competition}---to identify which component removes the deployment bottleneck and which preserves geometry under a tight codec budget.

\subsection{Experimental Protocol}

\paragraph{Hardware and codec settings.}
Main experiments were conducted on a desktop with a 13th Gen Intel Core i9-13900K CPU (3.00\,GHz), 32\,GB system RAM, and an NVIDIA RTX 4090 GPU with 24\,GB VRAM.
Codec evaluation uses FFmpeg 5.1 HEVC with the medium preset. Desktop packages use YUV444, while GalaxyXR packages use YUV420.
For AtlasLC, all experiments use a \(512\times512\) atlas with \(K=1\) slot per pixel (\(T=262{,}144\)), a \(64\times64\) local-competition grid, and a \(16\times16\) packing grid.

We report averages over the 320-object Objaverse evaluation
set~\cite{Deitke2023_Objaverse} and the 30-object
Stanford-ORB~\cite{Kuang2023_StanfordORB}, where each object is
evaluated on 48 and 10 held-out views, respectively.
Because our target is deployable object-centric 3DGS for XR, we
evaluate four aspects together: compactness (payload), image
fidelity (PSNR, SSIM, LPIPS), geometry (3D F1, depth RMSE), and
deployment efficiency (atlas preparation, encoding, decoding,
FPS, and total pipeline time).
AtlasLC uses the globally fixed input-adaptive budget in
Sec.~\ref{sec:adaptive_budget}. Its budget depends only on the
input Gaussian count and the fixed atlas capacity, without
rendered-quality or per-object validation feedback.
material.

\paragraph{Geometry evaluation.}
We report 3D F1 to measure bidirectional geometric consistency between the predicted point set \(P\) and sampled GT mesh surface points \(G\).
Specifically, precision is defined as the fraction of predicted points that lie within a distance threshold \(\tau\) of the GT surface, and recall is defined as the fraction of GT surface samples that are covered within \(\tau\) by the prediction.
We then report
\[
\mathrm{F1}_{3D}=
\frac{2\,\mathrm{Prec}_\tau\,\mathrm{Rec}_\tau}
{\mathrm{Prec}_\tau+\mathrm{Rec}_\tau}.
\]
Unless otherwise noted, we sample 30k points from the GT mesh surface, subsample the predicted point set to at most 30k points, and set \(\tau\) to 1\% of the GT bounding-box diagonal. Higher is better.

\paragraph{Depth evaluation.}
For depth RMSE, we render per-view depth maps from the Gaussian representation using RaDe-GS~\cite{Zhang2026_RaDeGS}.
Unless otherwise noted, the same rasterized depth formulation is applied to the decompressed Gaussian assets of all Gaussian-based methods so that depth error is measured under a consistent extraction rule across methods.

VANILLA denotes the uncompressed 3DGS reference~\cite{Kerbl2023_3DGS}.
PUP is used for the prune-only comparison, while PLAS~\cite{Morgenstern2024_SelfOrganizingGaussianGrids}, LGSCV~\cite{Yang2025_LGSCV}, UVGS~\cite{Rai2025_UVGS}, and FLEX*~(FlexGaussian~\cite{Tian2025_FlexGaussian}) are included in the full-pipeline comparison.
Map denotes layout, atlas, or UV-assignment construction; Encode denotes standard-codec packaging or a method's native compression; Decode denotes decompression and reconstruction into renderable Gaussian parameters; and Total is the sum of all applicable measured stages. FlexGaussian has Map = N/A because it does not construct a 2D atlas or UV map; its native parameter compression is counted under Encode and its dequantization/reconstruction under Decode.

\subsection{Main Comparison: Deployment-Aware Compression}

Table~\ref{tab:main_comparison} reports a representative
deployment operating point on the 320-object Objaverse evaluation
set, while Fig.~\ref{fig:rate_geometry} summarizes rate-dependent
PSNR, total pipeline time, and the precision and recall components
of 3D geometry preservation.
The main result is that \emph{heavy layout generation remains a real bottleneck for object-centric asset compression}.
AtlasLC largely removes this bottleneck and shifts the operating point toward lower payload, lower total pipeline time, and higher average 3D F1, with a small PSNR trade-off relative to these structured baselines.

Overall, AtlasLC provides a favorable deployment-aware operating point among the evaluated compressed baselines.
At this representative point, it obtains the lowest average payload (0.643\,MB), lowest decode time (108.1\,ms), highest FPS (135.8), highest average 3D F1 (0.865), and lowest total pipeline time (365.6\,ms) among the compressed methods, while remaining close to the strongest structured baselines in image-space fidelity.

The reduction in mapping overhead is especially significant.
Relative to PLAS, AtlasLC reduces atlas-preparation time 25.3$\times$ and shortens total time 5.54$\times$.
Relative to LGSCV, it reduces atlas-preparation time 13.3$\times$ and total time 3.37$\times$.
These results directly support our claim that preprocessing and map generation become a first-order bottleneck when scene-scale structured compression pipelines are applied to reusable object assets.

To avoid over-interpreting this representative operating point, we additionally perform matched-payload interpolation.
The common AtlasLC/PLAS/LGSCV range is 0.700--0.952\,MB.
UVGS has no valid payload overlap with AtlasLC in this sweep, so we do not make equal-rate claims against UVGS.
Over the matched range, AtlasLC trades 0.20--0.30\,dB lower PSNR for +0.005 3D F1 and 1.47--1.66\,s lower total time against PLAS, and 0.09--0.21\,dB lower PSNR for +0.007 3D F1 and 0.90--0.98\,s lower total time against LGSCV.

Paired bootstrap over Objaverse assets supports the geometry claim:
AtlasLC improves 3D F1 over PLAS by +0.0050, with a 95\% CI of [+0.0014,+0.0089], and over LGSCV by +0.0070, with a 95\% CI of [+0.0036,+0.0107].
LPIPS, depth RMSE, FPS, and decode-time differences are therefore reported as trends when their intervals cross zero.

FlexGaussian provides a cross-family comparison because it compresses Gaussian parameters rather than constructing a standard-codec atlas.
For this comparison, the matched point uses the AtlasLC 0.851\,MB operating point rather than the 0.643\,MB point in Table~\ref{tab:main_comparison}; at this point, AtlasLC improves PSNR by +1.326\,dB, 3D F1 by +0.014, decode time by -29.7\,ms, and total time by -108.7\,ms.

Table~\ref{tab:main_comparison_stanfordorb} further supports this trend on the object-only Stanford-ORB dataset.
While the photometric ranking is more mixed than on Objaverse, AtlasLC retains a favorable deployment profile, combining the lowest payload and total pipeline time with the highest FPS and 3D F1 among the evaluated compressed baselines.
This suggests that the deployment-oriented behavior of AtlasLC is not limited to virtual object assets, but also transfers to high-quality real-world object captures.

Qualitatively, Figs.~\ref{fig:rgb_compare} and \ref{fig:depth_compare} compare representative examples against VANILLA, PLAS, UVGS, and FLEX.
We use PLAS as the representative structured atlas baseline in the main paper because PLAS is slightly stronger quantitatively than LGSCV overall.
Full qualitative comparisons including LGSCV, across additional scenes and viewpoints, are provided in the supplementary material.

AtlasLC approaches VANILLA closely in geometry: the 3D F1 gap is only 0.005, while the payload is reduced from 22.223\,MB to 0.643\,MB.
Because AtlasLC is intended for deployable XR object assets, we additionally evaluate the client-side stage on a head-mounted device.
In this setting, mapping and HEVC encoding are performed on the server, and only the compressed package is transmitted to the HMD.
The device then decodes the received package and renders the decompressed Gaussian asset.
Accordingly, Table~\ref{tab:hmd_only} isolates the HMD-side deployment behavior and excludes server-side mapping and encoding cost, which are already reported in the desktop pipeline measurements.

\begin{table}[t]
\centering
\caption{On-device HMD evaluation after server-side packaging. Tx@5Mbps denotes the estimated one-shot server-to-HMD transfer time under a fixed 5~Mbps downlink, computed from payload.}
\label{tab:hmd_only}
\small
\setlength{\tabcolsep}{3.5pt}
\resizebox{\columnwidth}{!}{%
\begin{tabular}{lcccccc}
\toprule
Method & Payload (MB)$\downarrow$ & Tx@5Mbps (s)$\downarrow$ & PSNR$\uparrow$ & 3D F1$\uparrow$ & Decode (ms)$\downarrow$ & FPS$\uparrow$ \\
\midrule
PLAS    & 0.496 & 0.79 & \textbf{24.480} & 0.644 & 143.276 & 22.33 \\
LGSCV   & 0.530 & 0.85 & 24.234 & 0.639 & 430.732 & 19.91 \\
UVGS    & 1.031 & 1.65 & 20.390 & 0.612 & 145.406 & 30.82 \\
AtlasLC & \textbf{0.453} & \textbf{0.72} & 24.131 & \textbf{0.656} & \textbf{133.044} & \textbf{45.69} \\
\bottomrule
\end{tabular}%
}
\end{table}

\subsection{On-Device HMD Evaluation}
The main result is that the deployment advantage of AtlasLC carries over to the actual device (GalaxyXR) setting. We exclude FLEX*~\cite{Tian2025_FlexGaussian} since it is incompatible with the standard codec used on GalaxyXR.
AtlasLC achieves the fastest on-device decode time and the highest average FPS, while also achieving the highest average 3D F1 at the evaluated
operating point among the compressed baselines.
Although PLAS attains a slightly higher PSNR, AtlasLC remains competitive in image quality while providing a clearly stronger runtime operating point on the HMD.

Payload is especially important in this setting because server-to-device delivery time scales directly with the transmitted package size.
For interpretability, the measured HMD payload can be converted into an illustrative one-shot transfer proxy under a 5~Mbps downlink as
\[
t_{\mathrm{tx}} \approx \frac{8B}{5}\ \text{s},
\]
where \(t_{\mathrm{tx}}\) denotes the estimated server-to-HMD transmission time, \(B\) is the payload in MB, and the resulting value is expressed in seconds (\(\text{s}\)).
This quantity is a payload-derived proxy rather than a measured live networked-delivery result. Under this proxy, AtlasLC yields the shortest estimated server-to-HMD transmission time, reinforcing the same deployment trend observed in decode time and FPS.

\subsection{Why Local-Competition Pruning Helps}

We next isolate AtlasLC's pruning rule by comparing its \emph{local-competition pruning} against PUP in a prune-only setting at matched or nearly matched kept-Gaussian counts (Tab.~\ref{tab:pup_atlaslc_prune_only_30_90_compact}).
The survivor counts are exactly matched at 90\%, 80\%, 50\%, 40\%, and 30\% pruning, and differ by only one Gaussian at 70\% and 60\%, so performance differences reflect \emph{which} Gaussians are retained rather than how many.

The main pattern is clear: AtlasLC provides the largest advantage in the aggressive pruning regime, where survivor selection is most critical.
At 90\% pruning, it improves PSNR by 0.498\,dB, SSIM by 0.068, reduces LPIPS by 0.015, and increases 3D F1 by 0.053.
At 80\%, the gains are +0.592\,dB PSNR, +0.061 SSIM, -0.013 LPIPS, and +0.063 3D F1.
At 70\%, 60\%, 50\% AtlasLC still improves all four metrics, and at 50\% the result remains favorable overall.

At lighter pruning ratios, the effect becomes more selective.
At 40\% and 30\% pruning, PSNR and SSIM are nearly unchanged and LPIPS is slightly worse (+0.001 and +0.002), but 3D F1 still improves by 0.029 and 0.041.
This suggests that when the pruning budget is loose, both methods already preserve most photometric support, while the main remaining benefit of local competition is better preservation of geometry-relevant support.

Figure~\ref{fig:pup_vs_scotlc_pruning} is consistent with this interpretation.
PUP prunes splats more diffusely across the object support, whereas AtlasLC removes Gaussians in compact local groups.
This is the intended effect of local competition: neighboring splats that explain similar support compete against one another, so redundancy is removed \emph{within} a neighborhood rather than subtracted more globally from the object surface.
Under tight budgets, this yields clear gains in both image fidelity and geometry; under looser budgets, the advantage appears primarily in geometry preservation.

Taken together, Tab.~\ref{tab:pup_atlaslc_prune_only_30_90_compact} and Fig.~\ref{fig:pup_vs_scotlc_pruning} support the design of AtlasLC's pruning stage.
For object-centric assets, pruning should resolve redundancy at the local-support level rather than by a purely global removal rule.

\subsection{Ablation}

Table~\ref{tab:ablation_scot_lc_det_no_decode_outliers} ablates the two main design choices of AtlasLC while keeping the shared ordering backbone fixed.
The clearest result is that \emph{deterministic atlas packing} is the dominant systems component: it is the stage that turns a reasonable retained set into a compact, codec-efficient, and fast-to-deploy asset.

The most consequential comparison is \emph{LC pruning only} versus full \emph{AtlasLC}, which isolates deterministic packing.
Without deterministic packing, the pruned representation is not yet a practical codec asset:
payload grows from 0.643\,MB to 2.848\,MB, PSNR drops from 27.034 to 25.195, SSIM drops from 0.9301 to 0.8820, LPIPS worsens from 0.0732 to 0.0902, 3D F1 drops from 0.865 to 0.853, decode time rises from 108.1\,ms to 152.7\,ms, and total pipeline time rises from 365.6\,ms to 454.1\,ms.
Adding deterministic packing reduces payload by 4.43$\times$, improves PSNR by 1.839\,dB, improves 3D F1 by 0.012, lowers decode time by 44.6\,ms, and lowers total time by 88.5\,ms.
Deterministic packing is therefore not a cosmetic implementation detail; it is the mechanism that removes the mapping/remapping bottleneck while making the final asset compact and codec-friendly.

We then compare \emph{Deterministic packing only} with full \emph{AtlasLC} to isolate local-competition pruning.
Without LC pruning, the model reaches the strongest photometric scores among our variants:
27.314\,dB PSNR, 0.9393 SSIM, and 0.0725 LPIPS, together with the fastest atlas-preparation stage (51.7\,ms).
However, this higher-fidelity point comes with a larger payload (1.003\,MB vs.\ 0.643\,MB) and slightly weaker geometry (3D F1 0.864 vs.\ 0.865; depth RMSE 0.0366 vs.\ 0.0361).
Adding LC pruning reduces payload by 35.9\%, improves 3D F1 and depth RMSE, increases FPS from 129.3 to 135.8, and reduces decode time from 118.2\,ms to 108.1\,ms, at the cost of a modest drop in raw image-space fidelity.

The ablation reveals a clean division of labor.
The shared ordering backbone provides a lightweight coordinate system reused by all variants.
Deterministic packing decides \emph{how} retained support is laid out for compression and reconstruction, and is essential for payload, decoding efficiency, and total pipeline time.
Local-competition pruning decides \emph{which} support survives under a tight budget, improving compactness and geometry.

\section{Limitations and Future Work}

AtlasLC is evaluated using deployment-oriented metrics spanning
payload, atlas-preparation time, decode latency, rendering FPS,
3D F1, and depth RMSE. These metrics provide useful proxies for
asset compactness, preparation and runtime efficiency, and the
preservation of geometry and depth cues, but they do not directly
measure downstream user performance or perceived interaction
quality in live XR experiences. Accordingly, our claims concern
deployment-aware asset quality rather than demonstrated
interaction benefits. In particular, we do not yet test whether
the observed geometry preservation and latency reductions
translate into more accurate occlusion judgments, more stable
object boundaries during interaction, or higher accuracy in
selection and manipulation tasks. A natural next step is therefore
to pair AtlasLC with task-level user studies that measure startup
responsiveness, occlusion correctness, and interaction performance
under matched payload budgets.

A second limitation concerns scope. AtlasLC targets released,
static, object-centric Gaussian assets in a source-free setting
and is evaluated with a fixed codec-friendly atlas pipeline. This
makes the method practical for reusable XR object libraries, but
the present study does not yet address dynamic, articulated, or
deformable assets, scene-scale multi-object captures, or
alternative device- and codec-specific packaging strategies.
More broadly, because AtlasLC is intentionally post hoc and
training-free, it does not use original multiview images or
per-asset optimization that could further improve absolute
rate--distortion performance. Future work could explore
device-adaptive packing and streaming, as well as a source-assisted
authoring-to-deployment pipeline. Given a single image or a small set of source views, efficient segmentation models could isolate the target object \cite{Zhao2023_FastSAM, Kim2025_SemanticFastSAM}. A single-image Gaussian
generator such as LGM~\cite{Tang2024_LGM} could then create an
object-centric Gaussian asset, after which AtlasLC could perform
the final compression and standard-codec packaging for
transmission, caching, and XR deployment. This source-assisted
extension is complementary to, and outside, the source-free
setting evaluated in this work.

A further limitation is the globally fixed opacity-only survivor
score. The observed failure cases are predominantly photometric
and concentrate in thin, transparent, or high-anisotropy
silhouette-support regions, where visually important low-opacity
splats can be under-ranked relative to nearby opaque support in
the same local-competition cell. Because the final counts of
packing collisions, dropped retained Gaussians, and unplaced
retained Gaussians are all zero, these observations suggest that
the artifacts primarily arise from the limits of source-free local
selection rather than atlas placement. Additional examples and
the full failure taxonomy are provided in the supplementary
material. Future work could investigate lightweight
multi-attribute survivor scores that combine opacity with scale or
anisotropy while preserving source-free operation; source-assisted
variants could additionally exploit boundary or semantic cues. 

\section{Conclusion}

We presented \textbf{AtlasLC}, a source-free, training-free compression pipeline for object-centric 3D Gaussian Splatting aimed at deployable XR assets.
AtlasLC combines \emph{local-competition pruning} and \emph{deterministic atlas packing} over a lightweight shared ordering backbone, removing the heavy mapping/remapping overhead of prior structured pipelines while preserving object-wide foreground support.
Across the evaluated representative operating points, AtlasLC provides a favorable combination of payload, preparation time, decode latency, FPS, and 3D F1 among the compressed baselines.
These results support a simple conclusion: for object-centric 3DGS in XR, the right target is not bitrate alone, but a \emph{deployment-aware operating point} that jointly balances compactness, preparation cost, decoding efficiency, and geometry preservation.
We hope AtlasLC helps make scalable libraries of reusable radiance-field object assets more practical for XR and motivates future work on interaction-aware, device-aware, and codec-friendly deployment of 3DGS assets.

\acknowledgments{%
This work was supported by the Institute of Information \&
Communications Technology Planning \& Evaluation (IITP) grant
funded by the Korea government (MSIT)
(RS-2019-II191270, WISE AR UI/UX Platform Development for
Smartglasses).
This research was supported by the Institute of Information \&
Communications Technology Planning \& Evaluation (IITP) under
the virtual convergence support program to nurture the best talents
(IITP-2022(2026)-RS-2022-00156435) grant funded by the Korea
government (MSIT).
This work was supported by Institute of Information \& Communications Technology Planning \& Evaluation (IITP) grant funded by the Korea government (MSIT) (RS-2026-25523396, Core Technology Development for Immersive Content).
}

\clearpage

\bibliographystyle{abbrv-doi-hyperref-narrow}
\bibliography{template}

\end{document}